\documentclass[10pt, twocolumn, conference]{IEEEtran}

\usepackage{graphicx}
\usepackage{amsmath, amssymb}
\usepackage{citesort}

\title{Network Coding over a Noisy Relay : a Belief Propagation Approach}
\author{\authorblockN{Sichao Yang and Ralf Koetter}
\authorblockA{Coordinated Science Laboratory \\ Department of Electrical and
Computer Engineering \\ University of Illinois at Urbana-Champaign \\ \{syang8,
koetter\}@uiuc.edu}
}

\begin{document}

\maketitle

\begin{abstract}
In recent years, network coding has been investigated as a method to obtain
improvements in wireless networks. A typical assumption of previous work is that
relay nodes performing network coding can decode the messages from sources
perfectly. On a simple relay network, we design a scheme to obtain network
coding gain even when the relay node cannot perfectly decode its received
messages. In our scheme, the operation at the relay node resembles message
passing in belief propagation, sending the logarithm likelihood ratio (LLR) of
the network coded message to the destination. Simulation results demonstrate the
gain obtained over different channel conditions. The goal of this paper is not
to give a theoretical result, but to point to possible interaction of network
coding with user cooperation in noisy scenario. The extrinsic information
transfer (EXIT) chart is shown to be a useful engineering tool to analyze the
performance of joint channel coding and network coding in the network.
\end{abstract}

\section{Introduction}
Since the seminal paper by Ahlswede et al. \cite{AhlswedeNetwork00}, network
coding has been investigated as a potential tool for the design of communication
networks in order to let the data transmission rate approach the
capacity limit. Some recent work
\cite{Hausl05Iterative,ChenWireless06,KattiXORs06} studies the application of
network coding to wireless networks as a way for providing users with
cooperative diversity. All these papers show that network coding does have
practical benefits and can substantially improve wireless throughput.

\begin{figure}[ht]
 \centering
 \includegraphics[width=7cm]{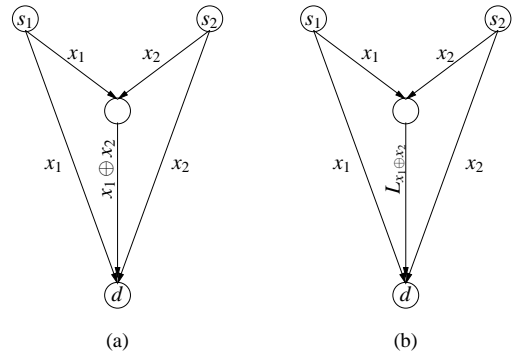}
 \caption{Network coding over a noisy relay by belief propagation}
 \label{fig.relaynet}
\end{figure}

Our paper was motivated by a simple relay network presented in
\cite{Hausl05Iterative}. The structure of the relay network from
\cite{Hausl05Iterative} is shown in Figure \ref{fig.relaynet}.(a). There are
two sources $s_1, s_2$ and one destination $d$. Both sources broadcast their
coded messages to the relay and destination. The relay helps the transmission
by sending its observation of the sources to the destination. The different
point-to-point channels are assumed to be Gaussian and non-interfering in
\cite{Hausl05Iterative}. The crucial assumption in \cite{Hausl05Iterative} is
that the relay node is assumed to be able to decode the messages from both the
sources reliably.

We note that even in a single source and single relay network, the accurate
capacity is still unknown.  In \cite{Hausl05Iterative}, the authors evaluate
the performance of the following scheme in the relay network shown in Figure
\ref{fig.relaynet}.(a): LDPC codes are applied in all point-to-point channels.
Let $x_1$ and $x_2$ be the bits sent by the sources. Let $\varphi(x): \{0,1\}
\rightarrow \{+1, -1\}$ be the standard BPSK modulation map .  The relay
decodes the signal from the source perfectly and it sends $x = \varphi(x_1
\oplus x_2)$ to the destination. Then, the channel can avail of three
observations $y_1 = \varphi(x_1) + n_1, y_2 = \varphi(x_2) + n_2, y =
\varphi(x_1 \oplus x_2) + n_r$ from three independent channels. The
destination jointly decodes $x_1$ and $x_2$ by these three observations. The
authors in \cite{Hausl05Iterative} compare their scheme to a reference scheme,
in which the relay spends half of her effort for helping $s_1$ and half for
helping $s_2$ in a time division manner. The scheme using network coding shows
a significant improvement.

In \cite{Hausl05Iterative}, as well as other previous work such as
\cite{ChenWireless06,KattiXORs06}, a fundamental assumption is that the relay
node is able to decode the source messages reliably. This assumption limits
their investigation to the cases that the channels from sources to the relay
have good quality and the channel codes applied are strong enough. However, in
reality, the relay may be far away from the sources so that the channels from
sources to the relay are subject to high noise or severe fading. % Furthermore,
% the nodes in a wireless network may have power or computational complexity
% limits so that the application of strong channel codes is questionable.

In this paper, we aim to investigate how network coding gain may be achieved
in a wireless networks even when the transmission to the relay can not be
recovered perfectly. The basic idea of our scheme is the following: Instead of
decoding the messages from the sources, the relay node mimicks a message
passing belief propagation setup, transmitting the logarithm likelihood ratio
(LLR) of the network coded message to the destination as shown in Figure
\ref{fig.relaynet}.(b). We consider this approach a step towards reconciling
network coding and user cooperation in noisy environment and show the network
throughput improvement in our scheme.  Moreover, we hope that our work
provides initiative in rethinking the channel code design rules for the
cooperation in wireless networks.

The paper is organized as follows. In Section \ref{sec.sys}, the system model
and our scheme is introduced in details. In Section \ref{sec.simulation}, the
numerical results are presented. In Section \ref{sec.exit}, the EXIT chart
over the system is studied for performance analysis. In Section
\ref{sec.conclusion}, conclusions are made and some future directions are
discussed.

\section{System Model}
\label{sec.sys}
In this section, we introduce the system model in details. We study the relay
network shown in Figure \ref{fig.relaynet}, which has the identical topology
as the network studied in \cite{Hausl05Iterative}. Two sources $s_1$ and $s_2$
are independent binary random sources with equal probability for $0$ and $1$.
All the channels are Gaussian channels. All point-to-point channels are
interference-free, so that the relay is able to receive a network coded
message in one time slot and the destination is able to run an iterative
decoding algorithm synchronously. We further assume that the network is
symmetric: the two channels from the sources to the relay have the same
quality and the two channels from the sources to the destination have the same
quality.

We use the following notations for the noise and signal-to-noise ratio (SNR)
on different channels:
\begin{center}
\begin{tabular}{cl}
$N_{\mbox{sd}}$, $\mbox{SNR}_{\mbox{sd}}$: & Noise and SNR on the channel \\ &
from the sources to the destination. \\
$N_{\mbox{sr}}$, $\mbox{SNR}_{\mbox{sr}}$: & Noise and SNR on the channel \\ &
from the sources to the relay. \\
$N_{\mbox{rd}}$, $\mbox{SNR}_{\mbox{rd}}$: & Noise and SNR on the channel \\ &
from the relay to the destination.
\end{tabular}
\end{center}

\begin{figure}[ht]
 \centering
 \includegraphics[width=8cm]{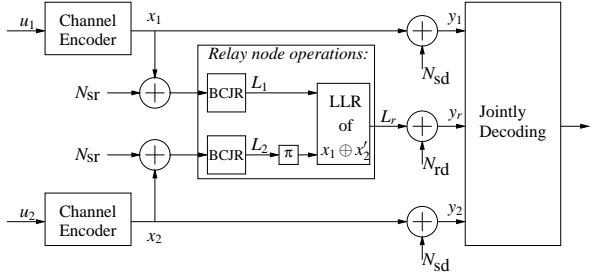}
\caption{Block diagram of  the system}
\label{fig.system}
\end{figure}

A systematic view of our scheme is shown in Figure \ref{fig.system}.

We use convolutional codes as channel codes, which is simple for the performance
analysis and illuminative when we demonstrate our ideas. The relay node
operation includes three steps:\\
\indent {\em Step 1}: The BCJR algorithms are run to derive the LLR for the
messages from each sources. In Figure \ref{fig.system}, $L_1$ and $L_2$ denote
the LLR of messages from $s_1$ and $s_2$ respectively.   \\
\indent {\em Step 2}: Permute LLRs of a codeword from  $s_2$, which decrease the
channel dependency of the three messages sent to the destination. \\
\indent {\em Step 3}: Calculate the LLR of the network coded message by
\begin{eqnarray*}
  L_r = \log{\left(\frac{e^{L_1}+e^{L_2}}{1+e^{L_1+L_2}}\right)}
\end{eqnarray*}

\begin{figure}[htbp]
 \centering
 \includegraphics[width=8cm]{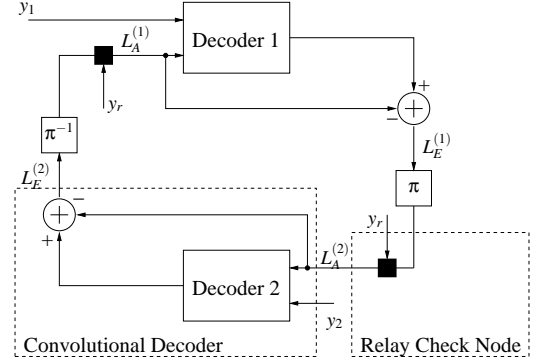}
 \caption{Iterative decoding}
 \label{fig.iter}
\end{figure}

The destination receives three messages: $y_1 = \varphi(x_1) + N_{\mbox{sd}}$,
$y_2 = \varphi(x_2) + N_{\mbox{sd}}$, and $y_r = L_r + N_{\mbox{rd}}$, where
$L_r$ is transmitted as analog value. The destination run an iterative
decoding algorithm based on these three observations. The iterative message
passing between two convolutional decoders are shown in Figure \ref{fig.iter}.

The additional operators between the convolutional decodes stand for the relay
check, of which the indicate function is
\begin{eqnarray}
T(x_1, x_2', x) = \left\{\begin{aligned}
                        & 1, &&\mbox{if}~x = x_1 \oplus x_2'. \\
            & 0, &&\mbox{otherwise}.
                       \end{aligned}\right.
\end{eqnarray}
Clearly, if the quality of relay channels is extremely bad, the decoder
appears to be two separate convolutional decoders. If the quality of relay
channel and the relay information $L_r$ are extremely
good, the decoder appears to be a simple turbo decoder. However, note that, in
contrast to classical turbo codes, all code bits in the two convolutionally
coded data streams are coupled because $x_1+x'_2$ is known. 

\section{Simulation results}
\label{sec.simulation}
\begin{figure}[htbp]
 \centering
 \includegraphics[width=5.5cm, angle=270]{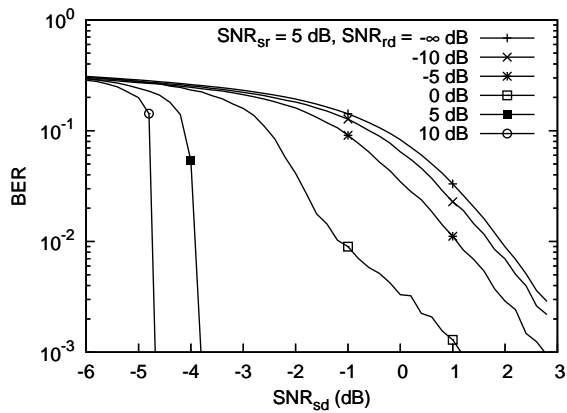}
 \caption{Bit error rate (BER) of the system with
   $\mbox{SNR}_{\mbox{sr}}=5\mbox{dB}$}
 \label{fig.5dB}
\end{figure}

\begin{figure}[htbp]
 \centering
 \includegraphics[width=5.5cm, angle=270]{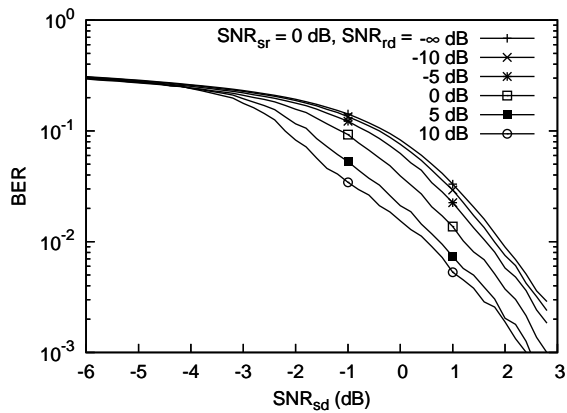}
 \caption{Bit error rate (BER) of the system with
   $\mbox{SNR}_{\mbox{sr}}=0\mbox{dB}$}
 \label{fig.0dB}
\end{figure}

We study the performance of the scheme presented in section \ref{sec.sys} by
simulation. In the simulation, we use the systematic rate $\frac{1}{2}$
recursive convolutional codes in the original Turbo codes \cite{BerrouNear93}
as the channel code, for which the code generator is
\begin{eqnarray}
G(D) = \left(1, \frac{1+D^4}{1+D+D^2+D^3+D^4}\right). \label{eq.generator}
\end{eqnarray}
We investigate the system performance under different channel conditions. The
simulation results are shown in Figure \ref{fig.5dB} and Figure \ref{fig.0dB}.
The figures show the bit error probability (BER) of the system in a
combination of different values of $\mbox{SNR}_{\mbox{sr}},
\mbox{SNR}_{\mbox{sd}}$ and $\mbox{SNR}_{\mbox{rd}}$. In Figure \ref{fig.5dB},
$\mbox{SNR}_{\mbox{sr}}$ is $5$dB. In Figure \ref{fig.0dB},
$\mbox{SNR}_{\mbox{sr}}$ is $0$dB. In both of the figures, the Y-axis is the
BER and X-axis is $\mbox{SNR}_{\mbox{sd}}$. The different curves in a figure
stand for the different $\mbox{SNR}_{\mbox{rd}}$, where
$\mbox{SNR}_{\mbox{rd}} = - \infty$ implies there is no relay. Obviously, if
the channel condition from sources to relay is better, more gain is obtained
through network coding.  In Figure \ref{fig.0dB}, when $\mbox{SNR}_{\mbox{sr}}$
is $0$dB, i.e, the channels are
not so good, there are still significant performance improvement.

\section{Performance Analysis by EXIT chart}
\label{sec.exit}
We briefly describe in this section a standard analysis tool called an
extrinsic information transfer (EXIT) chart to ease the selection of system
parameters, such as the channel code in the system. An EXIT chart, first
developed by Stephan ten Brink \cite{BrinkConvergence00}, is a technique to
aid the construction of good iteratively-decoded error-correcting codes (in
particular low-density parity-check (LDPC) codes and Turbo codes). EXIT charts
were built on the concept of extrinsic information developed in the Turbo
coding community. For the EXIT analysis, each component of the decoder (for
example a convolutional decoder of a Turbo code, the LDPC parity-check nodes
or the LDPC variable nodes) is modeled as a device mapping a sequence of
random variables $\mathbf{y}$ and $\mathbf{L}_i$ to a new sequence of random
variables $\mathbf{L}_o$, where $\mathbf{y}$ is the channel observation and
$\mathbf{L}_i$ and $\mathbf{L}_o$ are interpreted as LLRs for some random
source $X_i$ and $X_o$. For iterations between the components, the extrinsic
information is usually measured by mutual information $I(X_i, L_i)$ and
$I(X_o, L_o)$.

A key assumption in EXIT chart analysis is that the messages to and from a
component of the decoder can be described by a sequence of single numbers, the
a-priori information $\mathbf{L}_i$ and extrinsic information $\mathbf{L}_o$.
This is for example true when the sequence of observations $\mathbf{y}$ is
from a binary erasure channel. Otherwise, a crucial assumption in ten Brink's
analysis is that the sequence of information is reasonably approximated by
observations $\mathbf{y}$ from a Gaussian channel.

In this paper, the decoder has four different components as shown in Figure
\ref{fig.iter}: two convolutional decoders and two relay check nodes. In the
following discussion, we study the EXIT charts of the two components marked in
Figure \ref{fig.iter}. The EXIT charts of the other two components are the
same by symmetry.

Here we assume that both $y_r$, the observation of the relay check node, and
$y_2$ the observation of the convolutional decoder are from Gaussian channels.
That is,
\begin{eqnarray*}
&& y_2 = \varphi(x_2) + N_{\mbox{sd}} \\
&& y_r = \varphi(x_1 \oplus x_2) + N_{\mbox{r}}
\end{eqnarray*}
where $N_{\mbox{sd}}$ is the actually noise in the channels from the sources
to the destination and $N_r$ is assumed to be an approximation of the noise on
a concatenation of the channel from a source to the relay and the channel from
the relay to the destination. We let $X^{(1)}$ and $X^{(2)}$ denote the random
sources from $s_1$ and $s_2$.

\begin{figure}[htbp]
 \centering
 \includegraphics[width=6cm, angle=270]{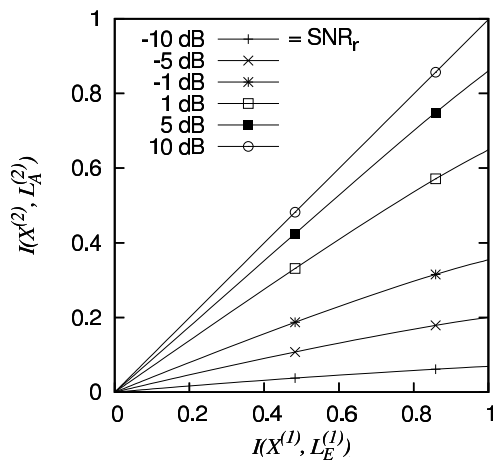}
 \caption{EXIT charts of the relay check node}
 \label{fig.EXITrelay}
\end{figure}

The EXIT chart of the relay node is shown in Figure \ref{fig.EXITrelay}.
Clearly, if the channel condition on the relay is bad, the information from
the other decoder barely pass through. However, if the channel condition on
the relay is good enough, all the information from the other decoder passes
through.

\begin{figure}[htbp]
 \centering
 \includegraphics[width=6cm, angle=270]{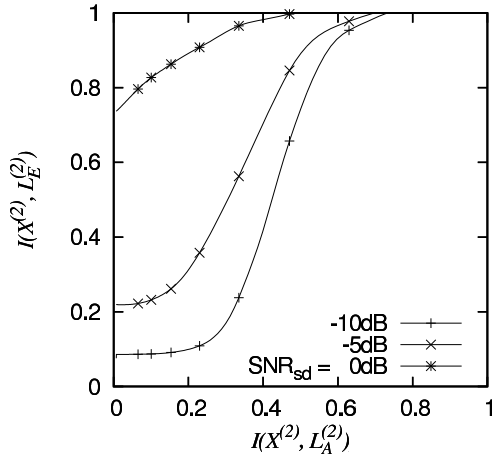}
 \caption{EXIT charts of the convolutional decoder}
 \label{fig.EXITorigin}
\end{figure}

The EXIT chart of convolutional decoders were extensively studied in many
papers. For illustration of our way in analyzing the system, we draw the EXIT
chart of the convolution decode with generator \eqref{eq.generator} in Figure
\ref{fig.EXITorigin}.

\begin{figure}[htbp]
 \centering
 \includegraphics[width=6.5cm]{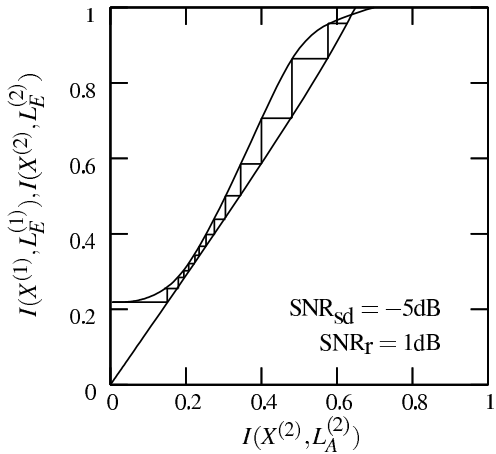}
 \caption{EXIT Charts and decoding trajectory of the system}
 \label{fig.EXITsystem}
\end{figure}

We analyze the system performance by the iteration of extrinsic information.
As shown in Figure \ref{fig.iter}, the extrinsic information of decoder $1$ is
$I(X^{(1)}, L_E^{(1)})$. After permutation, it changes to be the input of the
relay check node. The output of the relay check node is $I(X^{(2)},
L_A^{(2)})$, which is the a-priori information of decoder 2. After the
decoding process, the extrinsic information of decoder $2$, $I(X^{(2)},
L_E^{(2)})$, is derived. In Figure \ref{fig.EXITsystem}, the iteration of the
extrinsic information is shown in the case that $\mbox{SNR}_{\mbox{sd}} =
-5\mbox{dB}$ and $\mbox{SNR}_{\mbox{r}} = 1 \mbox{dB}$.

For a successful decoding, there must be a clear path between the curves so
that iterative decoding can proceed from $0$ bit of extrinsic information to
$1$ bit of extrinsic information. To make an optimal code, the two transfer
curves need to lie close to each other. This observation is supported by the
theoretical result that for capacity to be reached for a code over a
binary-erasure channel there must be no gap between the curves and also by the
insight that a large number of iterations are required for information to be
spread throughout all bits of a code. As shown in Figure \ref{fig.EXITsystem},
the iteration stops at a point with almost $1$ bit extrinsic information and
two curves are very close. Therefore, in case that $\mbox{SNR}_{\mbox{sd}} =
-5\mbox{dB}$ and $\mbox{SNR}_{\mbox{r}} = 1 \mbox{dB}$, the convolutional code
with generator \eqref{eq.generator} is good channel code for such channels.
However, if either $\mbox{SNR}_{\mbox{sd}}$ or $\mbox{SNR}_{\mbox{r}}$
decreased, the two curves will intersect at the middle of the chart. Channel
codes stronger than the codes in \eqref{eq.generator} are required for
reliable communication between the sources and the destination.

\section{Conclusions and future works}
\label{sec.conclusion}
In the paper, we investigate the users' network coding gain when the relay is
noisy in a simple relay network. We use EXIT charts in studying the
performance of joint channel coding and network coding schemes.

Wireless networks are subject to various crucial physical limits such as
channel capacity, power, and chip speed, The ultimate goal is to build theory
background and design engineering tools in finding optimal channel codes, i.e.
codes with lower complexity and low error probability. By taking advantage of
cooperative diversity gain such as network coding, some weak channel codes can
be found for reliable communication in a wireless network though they may work
outside of a point-to-point channel capacity limit in the network.

Some problems are interesting to explore in the future. In the simple relay
network presented in this paper, if the channel conditions from two sources to
the relay and the destination are different, it's not always better to do
network coding than just to decode forward or amplify forward the source with
better channel condition. It's important to design schemes which adapt to the
channel conditions. Furthermore, we design our scheme in a tentative way by
assuming that the signals from the relay to the destination are analog. It
will be interesting to investigate how the quantization of the signals affects
the performance system by rate-distortion theory.

\section{Acknowledgement}
The authors gratefully acknowledge the useful discussion with Lei Ying and
Prof. Bruce Hajek, University of Illinois at Urbana-Champaign.

\end{document}